\documentclass[structabstract]{aa}  
\usepackage[varg]{txfonts}
\usepackage{longtable,lscape}
\usepackage{amssymb}
\usepackage{graphicx}
\usepackage{pdfpages}
\begin{document}
   \title{Distribution and evolution of galaxy groups \\in the Ursa Major supercluster}


   \author{M. O. Krause
          \inst{1}, A.L.B. Ribeiro\inst{1} \and P.A.A. Lopes\inst{2}}
        

   \offprints{A.L.B. Ribeiro}

   \institute{Laborat\'orio de Astrof\'{\i}sica Te\'orica e Observacional\\ 
              Departamento de Ci\^encias Exatas
e Tecnol\'ogicas\\ Universidade Estadual de Santa Cruz -- 45650-000, Ilh\'eus-BA, Brazil\\
              \email{krausefisico@ig.com.br}\\
		\email{albr@uesc.br}
              \and Observat\'orio do Valongo, Universidade Federal do Rio de Janeiro -- 20080-090, Rio de Janeiro-RJ, Brazil\\
\email{paal05@gmail.com}}


 
  \abstract
{We study an SDSS sample of galaxies within $\sim$50 Mpc of the nominal
center of the Ursa Major supercluster.}{Our aim is to study galaxy distribution around groups in the supercluster and the link between the distribution of relaxed and nonrelaxed galaxy systems with respect to
the supercluster environment.} {Using the FoF algorithm, 40 galaxy groups were identified in this region. 
Velocity distributions for these groups were
studied after applying a 3$\sigma$-clipping routine for outlier removal. We classified the systems according to
the results of normality and substructure tests applied to member galaxies. Then, we studied
the relative distribution of relaxed and nonrelaxed systems across the supercluster.}{
We find that 68\% of galaxy groups are Gaussian and that all the
non-Gaussian systems have substructures and probably correspond to multimodal systems in redshift space. 
We also find that the Gaussian systems inhabit the
denser regions of the supercluster, with higher 
densities of both red and blue galaxies within 2.5 $h^{-1}$Mpc, and have smaller group-group pairwise separations.
}
{Our results suggest a spatial segregation of dynamical states, where
relaxed systems may have formed and evolved earlier and faster around high-density peaks, while
nonrelaxed systems may be growing slower on the peripheries of lower density peaks.
In this picture, galaxy clustering seems to be prompting a continuous
internal evolution in the supercluster, with several groups collapsing into the more evolved and contracted
regions.}

 
   \maketitle
%

\section{Introduction} 
Groups of galaxies contain most of galaxies in the Universe and are the link
between individual galaxies and large-scale structures
(e.g., Huchra \& Geller 1982; Geller \& Huchra 1983; Nolthenius \& White 1987; Ramella et al. 1989). 
The dissipationless evolution of these systems is dominated by gravity.
Interactions over a relaxation time tend to distribute the velocities of the galaxy members
into a Gaussian distribution (e.g. Bird \& Beers 1993).
Although the theoretical  line-of-sight ($los$) velocity distribution expected in galaxy systems is not 
exactly Gaussian (e.g. Merritt 1987; Kazantzidis et al. 2004),
phenomenological evidence has been suggesting for a long time that normality can
be assumed for systems in dynamical equilibrium (e.g. Yahil \& Vidal 1977). 

Several works have shown that the analysis of velocity distributions 
can separate galaxy groups into relaxed or nonrelaxed systems (e.g. Hou et al. 2009; Ribeiro et al. 2010, 2011).
Actually, several important properties of groups can be studied from this perspective.
For instance, Hou et al. (2009,2012) find rising velocity dispersion profiles for non-Gaussian (NG) groups. They have also shown that the majority of NG groups have substructures. Ribeiro et al. (2010) found 
that galaxies are redder and brighter in Gaussian (G) groups out to 4${\rm R_{200}}$. Martinez \& Zandivarez (2011) have
studied the luminosity functions of G and NG groups, concluding that G groups have
a brighter characteristic magnitude. Einasto et al. (2012a, 2012b) studied velocity distributions to
find signatures of multimodality in SDSS clusters.

The NG nature of the velocity distribution of NG groups must
be caused by recent group-group mergers. At the same time, NG groups
could be contaminated by chance projections. It is therefore important
to establish the real nature of NG groups and how they are linked to the environment. We study
an SDSS (Sloan Digital Sky Survey) sample of galaxies within $\sim$50 Mpc of the nominal
center of the Ursa Major Supercluster aiming to establish a relation
between large-scale environment and the evolution of galaxy groups. 
After identifying groups in the supercluster using the FoF (Friends-of-Friends) algorithm, 
we split up the sample into relaxed and nonrelaxed systems according to the
velocity distribution of member galaxies. Then, we investigate the distribution
of galaxy groups in the supercluster with respect to their dynamical state.
The paper is organized 
as follows. Section 2 presents the data and methods we have used, Section 3 presents our analysis, 
and Section 4 briefly summarizes the main conclusions of the work.
 
\section{Data}

In this work, we have used galaxies from the SDSS-DR6 (Adelman-McCarthy et al. 2008) within $\sim$50 Mpc around the coordinates of the
cluster Abell 1383 (
=177.46$^\circ$, Dec=54.62$^\circ$, and $z$=0.059) taken here as the nominal 
(and arbitrary) "center" of the Ursa
Major supercluster (hereafter UMaS).
The redshifts are in the range 0.045$< z <$0.075. We have found 18,204 galaxies in this region
with apparent magnitude limit $m_{\rm r}=17.77$.
Clusters were identified using the FoF algorithm (Huchra \& Geller 1982), following the
choice of linking lengths by Berlind et al. (2006) [$b_\perp=0.14$ and $b_\parallel=0.75$]\footnote{The parameters
$b_\perp$ and $b_\parallel$ are the projected and line-of-sight linking lengths in units of the mean intergalaxy separation.}, which were optimized for SDSS.
The FoF algorithm with these linking lengths finds galaxy groups with $N_{\rm gal}\geq$10 that have an unbiased multiplicity function (see Berlind et al. 2006).
These settings led us to find 40 galaxy groups in the UMaS, with numbers
of galaxies in the range $15 \leq N_{\rm gal}\leq 94$.
Our catalog contains all but one of the 23 galaxy systems known in the region \footnote{The missing cluster is beyond our search radius.} (see Kopylova \& Kopylov 2009). 

Velocity distributions for the identified groups were
studied after applying a 3$\sigma$ clipping routine for removing outliers.
First, a virial analysis \footnote{In this work, cosmology is defined by $\Omega_m$ = 0.3, $\Omega_\lambda$ = 0.7, and $H_0 = 100~h~{\rm km~s^{-1}Mpc^{-1}}$.} is done following Carlberg et al. (1997), where
$R_{200}=\sqrt{3}\sigma/[10H(z)]$ and $M_{200}=3R_{200}\sigma^2/G$.
Then, we applied three well known normality tests to the cleaned velocity distributions --
the Anderson-Darling (AD), Shapiro-Wilk (SW) and Jarque-Bera (JB) tests -- to decide if
a galaxy system is Gaussian or NG. Our criterium is that at least two tests
should reject normality at the 95\% confidence level to be considered NG;
if not, it will be considered Gaussian. Finally, we applied two substructure tests
-- the $\beta$  (West et al. 1988) and $\Delta$ tests (Dressler \& Schectman 1988) -- to
decide if a galaxy system has substructures.
Our criterion is that a galaxy system is considered to have substructures if at least
one of the two tests rejects regularity with 95\% confidence level.
Results are presented in Table 1, where columns are (1) group identification;
(2) and (3) equatorial coordinates in degrees; (4) average redshift of members after
removing outliers; (5) velocity dispersion in km/s;  (6) $R_{200}$ radius in $h^{-1}$Mpc;
(7) logarithm of $M_{200}$ in $h^{-1}$$M_\odot$;  (8) total number of galaxies after outlier removal;
(9) results for  the NG nature
of the $los$ velocity distribution: 0 for Gaussian and 1 for NG distributions;
(10) results for substructure tests: 0 for regular and 1 for irregular systems;
(11) Abell cluster identification.

\setlength{\tabcolsep}{0.15em}
\begin{table}
\begin{center}
\footnotesize{
\begin{tabular}{r c  c  c c c c c c c c}
\hline
 \# & RA  & Dec  & $\bar{z}$ & $\sigma$ & $R_{200}$ & log$M_{200}$ & $N_{\rm gal}$& V & S & Abell Id\\
\hline
1 & 197.07 & 39.83 & 0.0705 & 342 & 0.57 & 13.67 & 50 &1 &  1 & -\\
2 & 178.32 & 44.10 & 0.0707 & 250 & 0.42 & 13.26 & 33 &0 &  0 & - \\
3 & 172.31 & 56.28 & 0.0573 & 252 & 0.42 & 13.27 & 26 &1 &  1 & A1291\\
4 & 176.69 & 55.77 & 0.0785 & 387 & 0.64 & 13.83 & 94 &1 &  1 &  - \\ 
5 & 180.98 & 52.41 & 0.0624 & 181 & 0.30 & 12.84 & 25 &0 &  0 &  A1452\\
6 & 156.56 & 47.82 & 0.0620 & 282 & 0.47 & 13.42 & 67 &1 &  1  & A1003\\
7 & 183.54 & 60.00 & 0.0602 & 212 & 0.36 & 13.05 & 34 &0 &  0  & A1507\\
8 & 174.76 & 57.45 & 0.0665 & 190 & 0.32 & 12.90 & 21 &0 & 0   & A1318 \\
9 & 149.66 & 53.86 & 0.0470 & 272 & 0.46 & 13.38 & 36 &1 &  1  & - \\
10 & 176.75 & 55.69 & 0.0514 & 233 & 0.39 & 13.17 & 33 & 0 &  0 & A1377 \\
11 & 157.64 & 53.33 & 0.0643 & 232 & 0.39 & 13.17 & 30 & 0 &  1 & -\\
12 & 168.39 & 54.79 & 0.0710 & 193 & 0.32 & 12.92 & 29 & 0 &  0 & -\\
13 & 168.49 & 57.03 & 0.0474 & 255 & 0.43 & 13.29 & 39 & 0 &  0 &  - \\
14 & 181.69 & 56.93 & 0.0644 & 206 & 0.35 & 13.01 & 30 & 0 &  0 &  A1436\\
15 & 171.99 & 67.09 & 0.0549 & 325 & 0.55 & 13.61 & 56 & 0 &  1 & A1279\\
16 & 197.27 & 54.60 & 0.0642 & 220 & 0.37 & 13.10 & 21 & 0 &  0 & -\\
17 & 182.60 & 45.44 & 0.0660 & 200 & 0.33 & 12.97 & 36 & 0 &  0 & -\\
18 & 161.10 & 38.14 & 0.0428 & 337 & 0.57 & 13.66 & 55 & 1 &  1 & -\\
19 & 195.58 & 60.72 & 0.0702 & 275 & 0.46 & 13.38 & 40 & 1 &  1 & -\\
20 & 181.29 & 54.42 & 0.0503 & 232 & 0.39 & 13.17 & 28 & 0 &  1 & -\\
21 & 168.63 & 49.42 & 0.0733 & 259 & 0.43 & 13.31 & 39 & 0 &  0 & -\\
22 & 177.15 & 54.66 & 0.0596 & 207 & 0.35 & 13.02 & 27 & 1 &  1 & A1383\\
23 & 172.46 & 36.36 & 0.0625 & 275 & 0.46 & 13.39 & 35 & 0 &  0 & -\\
24 & 180.45 & 38.30 & 0.0646 & 207 & 0.35 & 13.02 & 21 & 0 &  0 & -\\
25 & 188.61 & 49.57 & 0.0403 & 336 & 0.57 & 13.65 & 49 & 1 &  1 & -\\
26 & 181.58 & 42.69 & 0.0526 & 282 & 0.48 & 13.42 & 35 & 1 &  1 & A1461\\
27 & 178.54 & 49.15 & 0.0541 & 257 & 0.43 & 13.30 & 53 & 0 &  0 & -\\
28 & 168.14 & 41.07 & 0.0728 & 287 & 0.48 & 13.44 & 41 & 1 &  1 & -\\
29 & 189.56 & 42.00 & 0.0656 & 206 & 0.35 & 13.01 & 21 & 0 &  0 & -\\
30 & 184.97 & 59.79 & 0.0444 & 331 & 0.56 & 13.63 & 55 & 1 &  1 & A1534\\
31 & 160.71 & 45.22 & 0.0493 & 298 & 0.50 & 13.49 & 51 & 0 &  0 &  -\\
32 & 170.37 & 47.03 & 0.0534 & 182 & 0.31 & 12.85 & 18 & 0 &  1 &  -\\
33 & 159.19 & 56.65 & 0.0458 & 308 & 0.52 & 13.54 & 58 & 0 &  0 &  -\\
34 & 202.42 & 44.78 & 0.0613 & 251 & 0.42 & 13.27 & 44 & 1 &  1 &  -\\
35 & 161.15 & 57.57 & 0.0732 & 264 & 0.44 & 13.33 & 40 & 0 &  0 &  - \\
36 & 197.49 & 49.56 & 0.0554 & 237 & 0.40 & 13.19 & 35 & 0 &  0 &  -\\
37 & 170.34 & 41.94 & 0.0601 & 151 & 0.25 & 12.61 & 15 & 0 &  0 &  -\\
38 & 203.11 & 57.41 & 0.0528 & 281 & 0.47 & 13.42 & 41 & 0 &  0 &  -\\
39 & 172.47 & 54.30 & 0.0693 & 215 & 0.36 & 13.07 & 33 & 0 &  0 & A1270\\
40 & 166.68 & 43.83 & 0.0583 & 221 & 0.37 & 13.10 & 32 & 0 &  0 & A1169\\
\hline
\end{tabular}
}
\caption{Properties of 40 galaxy groups identified in the UMaS.
Columns 9 and 10 indicate results for normality and substructure
tests described in the text.} 
\end{center}
\end{table}

\begin{figure*}
\centering
\includegraphics{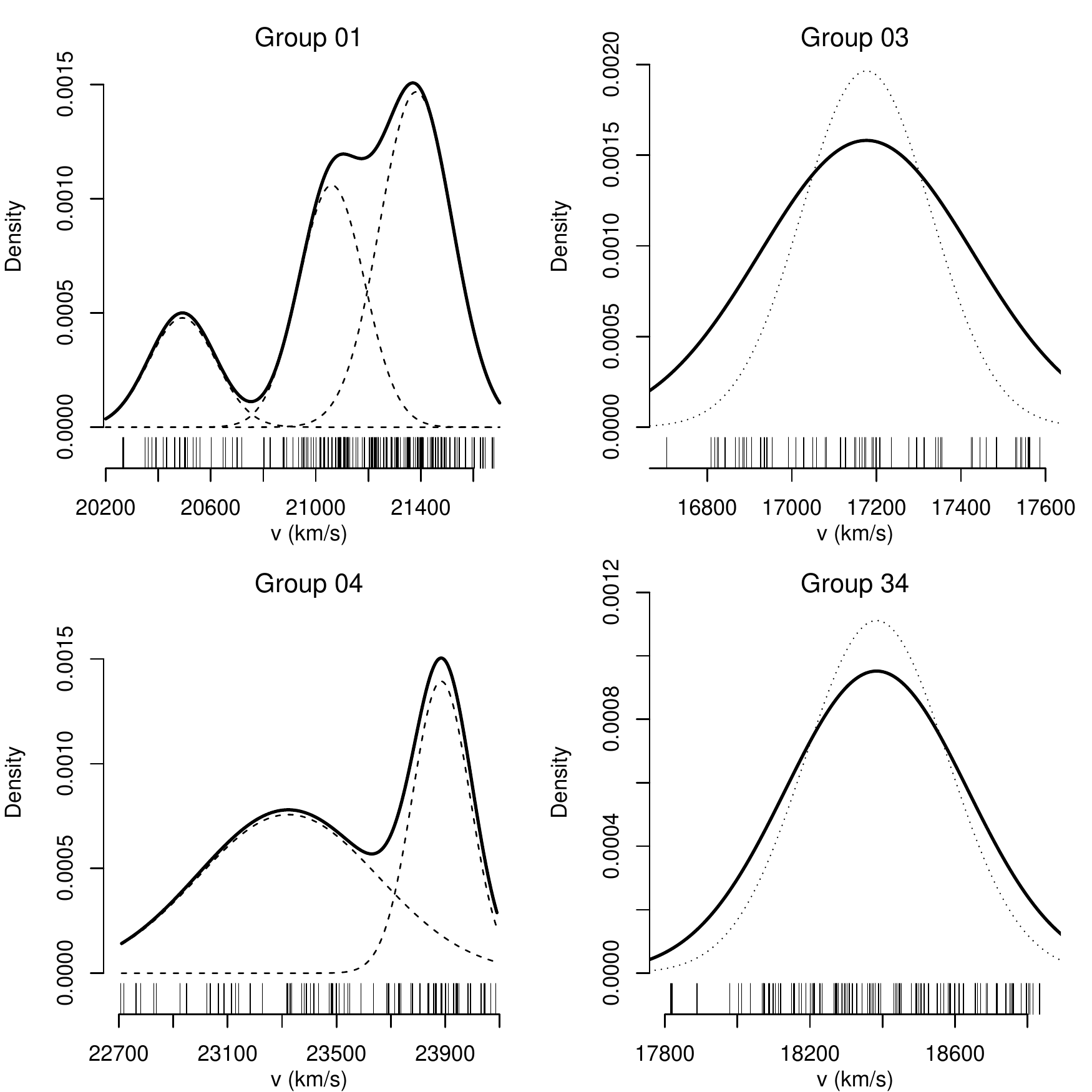}
\caption{Four examples of multimodality diagnostics for groups in the UMaS region. Vertical bars depict
individual galaxies in each group. The solid lines
indicate the best Gaussian mixture for for radial velocities distributions in km/s. 
The dashed lines show the individual components of the multimodal systems. The dotted lines
indicate the average Gaussian curve for 1,000 resamplings of groups 03 and 34, unimodal but non-Gaussian systems.}
\end{figure*}

\section{Analysis}
\subsection{Multimodality}

We find that 27 out of 40, i.e. 68\%, of galaxy groups 
are Gaussian and that all the NG systems have
substructures. Also, we find that four groups have substructures 
(as indicated by the $\beta$ test)
but have Gaussian velocity distributions according to our criteria. 
If we include them as nonrelaxed systems,
the total fraction of virialized groups would fall to 58\%. 
We take a conservative view to define nonrelaxed groups and assume that substructures are dynamically
significant only if they are also associated to non-Gaussianities in the velocity distribution of a
galaxy system.

\begin{figure*}
\centering
\includegraphics{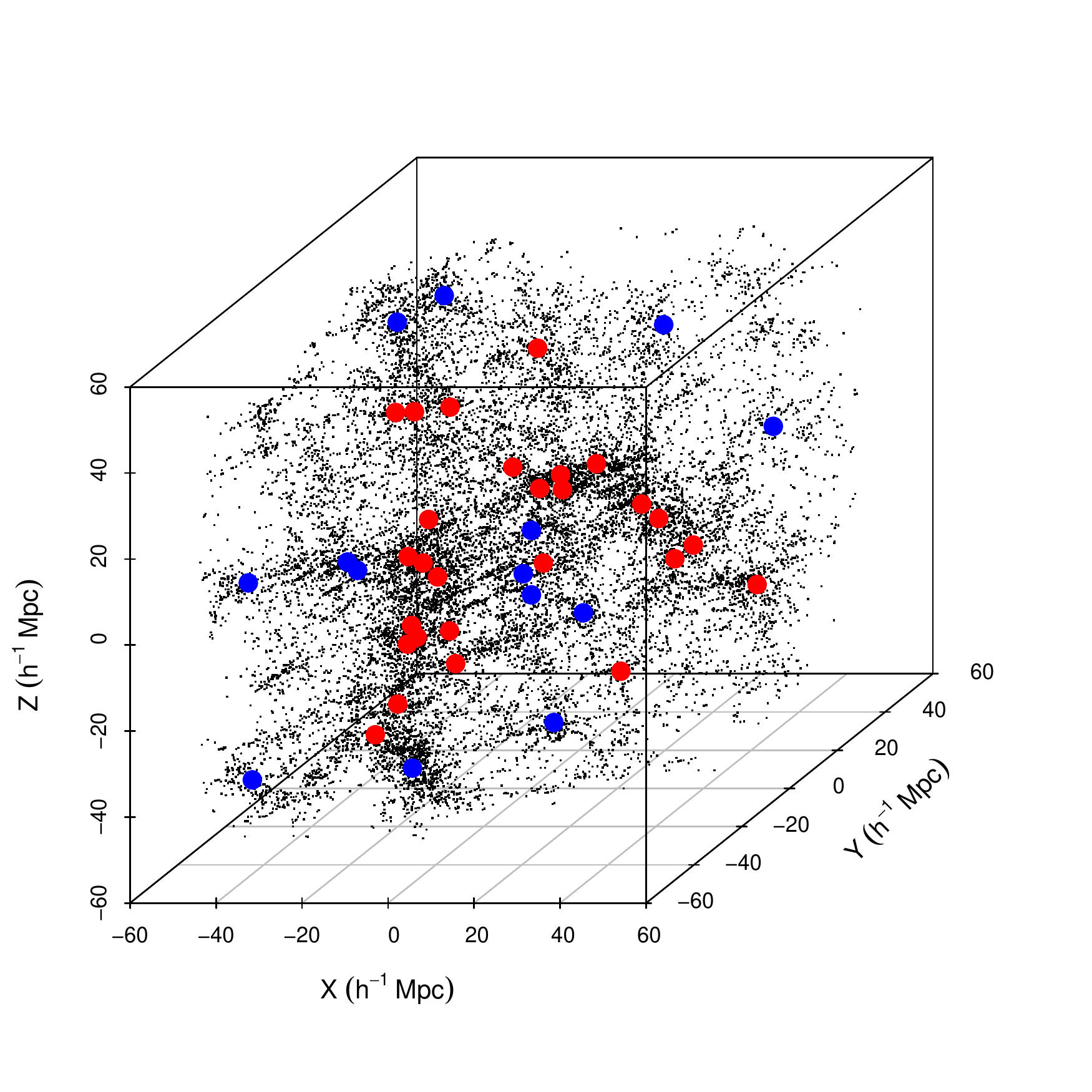}
\caption{Distribution of galaxies and galaxy systems in the UMaS in supergalactic coordinates.
Galaxies are in black; Gaussian groups  and non-Gaussian are in red and blue, respectively.}
\end{figure*}

Following Ribeiro et al. (2011),
we applied the  Kolmogorov-Smirnov (KS) and the Cram\'er-von Mises (CvM) two-sample tests on 
the G and NG subsamples. We find that NG velocity groups are signifcantly larger and more massive 
at the 95\% confidence level for both tests, with best fit relations:

\begin{itemize}
\item $\bar{R}_{200}^{NG}~\simeq ~(1.27\pm 0.22)~ \bar{R}_{200}^{G}$
\item[]  ~~~~~~~~~~~~~~~~~~~~~~~~~~~~~~~~~~~~~~~~~~~~~~~~~~~~~~~
\item $\bar{M}_{200}^{NG}~\simeq ~(2.00\pm 0.65)~ \bar{M}_{200}^{G}$ 
\end{itemize}

\noindent with [(p=0.0065 - KS), (p=0.0019 - CvM)] for the radius distributions, and 
[(p=0.0029 - KS) (p=0.0017 - CvM)] for the mass distributions.
This result is similar to the one found by Ribeiro et al. (2011),  a possible consequence of
applying the virial analysis to nonvirialized objects. Actually, we can assume the velocity distributions 
of NG groups as Gaussian mixtures with unknown
number of components, and use the Dirichlet process mixture (DPM) model to
study the velocity distributions (see Ribeiro et al. 2011 for details).
The number of components in the mixture is computed using the R language and environment 
(R Develoment Core Team) under the {\it dpmixsim} library
(da Silva 2009). We find the following results:

\begin{itemize}
\item Unimodal groups: 3, 9, 34
\item Bimodal groups: 4, 6, 18, 19, 22, 25, 26, 28, 30
\item Trimodal group: 1.
\end{itemize}

\begin{figure}[!h]
\resizebox{\hsize}{!}{
\includegraphics{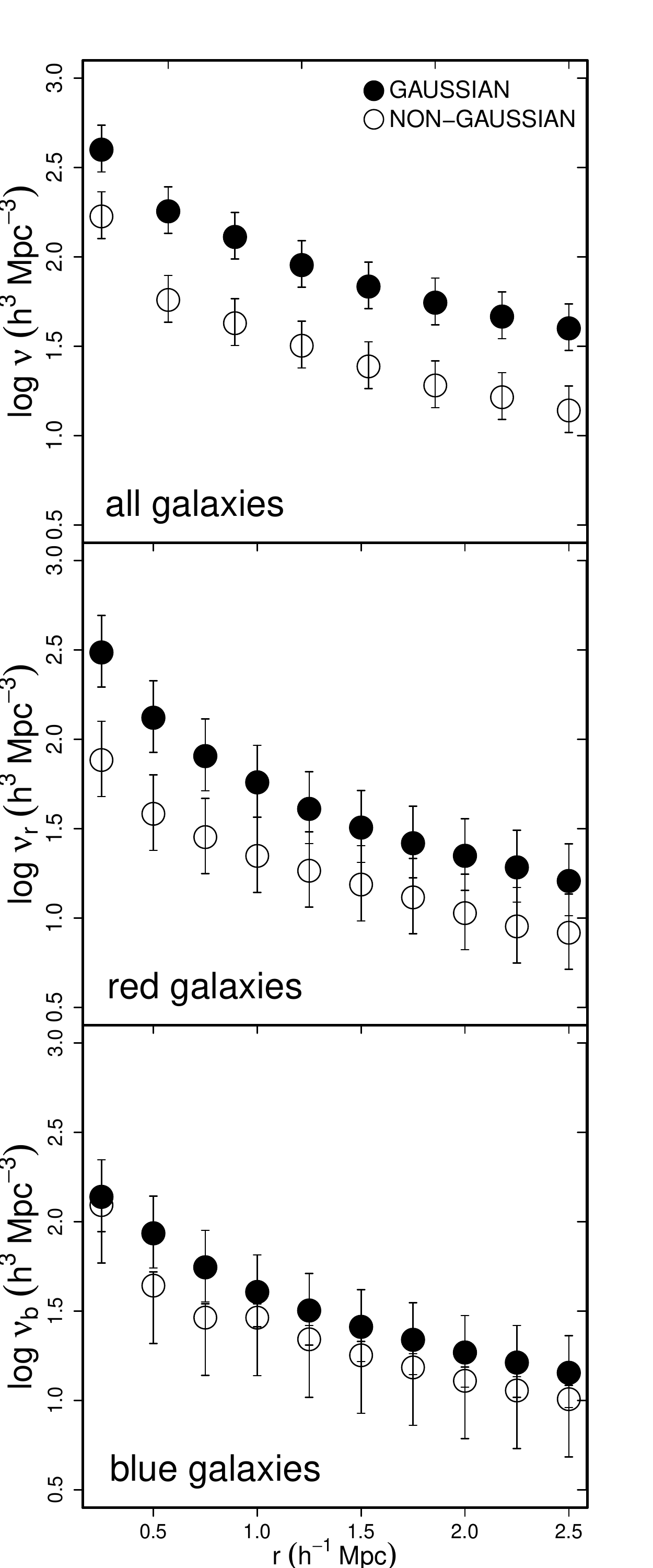}}
\caption{Galaxy number density profiles in redshift space. Filled and open circles denote G and NG systems,
respectively. Error bars are obtained from 1,000 bootstrap resamplings of galaxy groups. Top, middle and bottom plots refer to all, red, and blue galaxies, respectively.}
\end{figure}

 In Figure 1 we plot the velocity distributions for typical cases of NG groups.
Multimodality seems to be the prominent cause of non-Gaussianity
in galaxy group velocity distributions, corresponding to 10 out of 13, i.e. 77\% of the
NG cases in the UMaS. There are three unimodal systems (\#03, \#09, and \#34)
all having negative kurtosis excess ($k=-1.26$, $-0.87$, and $k=-1.05$, respectively) --
the mean values for Gaussian and NG groups are $k_{\rm G}=-0.34$ and $k_{\rm NG}=-1.16$
A negative kurtosis corresponds to a flattened and usually
multimodal distribution. However, if the separation between the modes
is too small, the DPM model is unable to recognize distinct components. 
In Fig. 1, we illustrate two cases of NG unimodal systems, also plotting
the Gaussian curve resulting from 1,000 resamplings keeping 90\% of the data.\footnote{For instance,
group \#03 has 26 members, thus each resampling provides a reduced group with 23 members.} For each
group we apply the normality tests mentioned in Section 2 to the new velocity distribution. The reduced samples
classified as normal distributions are used to compute the average Gaussian we plot in Figure 1.

\subsection{Supercluster Environment}

Now, we try to link the dynamical state of galaxy systems and the supercluster
environment. The spatial distribution of galaxy groups in the UMaS region is presented
in Figure 2 in supergalactic Cartesian coordinates\footnote{
The supergalactic coordinates are based on the preferred plane of the
Local Supercluster. After the coordinate transformation,
the Abell cluster A1383 was chosen to be at the
center of galaxy distribution.},
where we see galaxies and groups. From this figure,
it seems that G groups inhabit the denser regions of the supercluster, approximately
following the central arch-shaped structure in the plot. That could be
hinting at some environmental influence in the galaxy groups spatial distribution.

We have used two simple indicators to probe the possible relation between galaxy groups and 
the supercluster environment. 

\subsubsection{Distribution of galaxies around groups}

First, we built the stacked G and NG galaxy 
number density (in redshift space) up to 2.5 $h^{-1}$Mpc, around the groups' centers,
a distance corresponding to $\simeq 6\bar R_{200}$ for our sample.
Results are presented in Figure 3, where we see similar shapes
for the profiles around G and NG groups, but with densities systematically higher in the case of G groups. Error bars in this plot are obtained from a bootstrap technique with 1,000 resamplings applied to the galaxy groups. This result indicates that G systems present significantly richer surroundings than NG sytems, with
ratio of number densities of galaxies around G and NG groups following ${\rm log[\nu_{NG}}(r)/{\rm \nu_G}(r)]\simeq 0.35 \pm 0.10$ ($\nu$ is the galaxy number density).

We also classified galaxies into blue or red objects following the criterion
of Strateva et al. (2001) -- a red galaxy should have $u-r \geq 2.22$. This enables us to rebuild the galaxy density profiles for these two populations. The results for red galaxies are also presented in Figure 3 (the middle plot),
where we see that G groups have more red galaxies within 2.5 $h^{-1}$Mpc than NG groups
${\rm log[\nu_{NG}^r}(r)/{\rm \nu_G^r}(r)]\simeq 0.46 \pm 0.11$, but note that
the difference are smaller at larger radii, reflecting (i) the 
well known correlation between colors and environment -- the fraction of red galaxies
decreases toward the outer part of clusters (e.g. Balogh et al. 2000; De Propis et al. 2004), and (ii) the
quicker decline for G systems. At the same time, for blue galaxies the difference
between G and NG systems is smaller, but still statistically significant:
${\rm log[\nu_{NG}^b}(r)/{\rm \nu_G^b}(r)]\simeq 0.69 \pm 0.12$, see Figure 3
(on the bottom plot). These results show that G groups are denser and redder than
NG groups up to 2.5 $h^{-1}$Mpc.

\begin{figure*}
\centering
\includegraphics{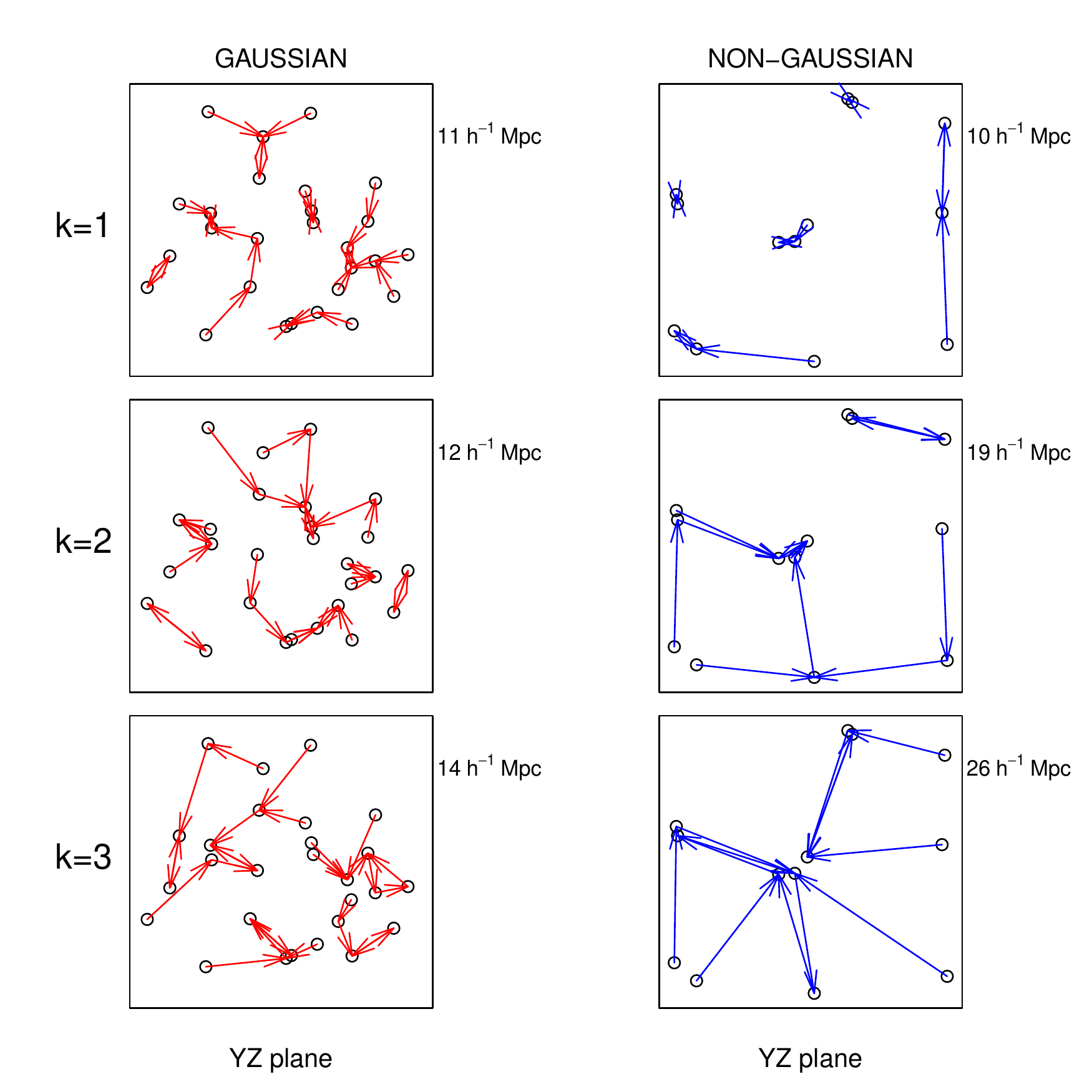}
\caption{Pairwise distances from each group to the first $(k=1)$, second $(k=2)$ and third $(k=3)$ nearest neighbors
in the supergalactic YZ projection. The arrows point to the $k$th nearest neighbor among groups of the same class.
The red and blue arrows indicate distances for G and NG systems, respectively. The mean distance between
groups appears on the right side of each box. The size of all boxes is 100 $\times$ 100 $h^{-1}$Mpc.}
\end{figure*}

\subsubsection{Distribution of groups through the UMaS}

Also, we studied the relative separation between groups in both G and NG subsets.
We computed the distance from each group to the first ($k=1$), second ($k=2$), and third ($k=3$) nearest groups
in the spatial distribution. In Figure 4, we illustrate our results for the supergalactic YZ projection\footnote{
We found approximately the same results for the XY and XZ projections}.
The mean pairwise distances of G groups show just a small variation
with $k$: 11 $h^{-1}$Mpc ($k=1$), 12 $h^{-1}$Mpc ($k=2$), and 14 $h^{-1}$Mpc ($k=3$);
while we see a wider variation for NG groups: 10 $h^{-1}$Mpc ($k=1$), 19 $h^{-1}$Mpc ($k=2$), and 26 $h^{-1}$Mpc ($k=3$).
This suggests that G groups are closer to each other than NG groups.

We estimated the significance of the pairwise separation test by scrambling
the identities of the G and NG groups 10,000 times, preserving the overall numbers of
27 Gaussian and 13 non-Gaussian. We calculated the number of replicas with more
extreme statistics than the input data (i.e. larger differences between G and NG mean pairwise separations).
For $k=1$, we found 4,356 cases, indicating that G and NG systems have the same distributions
with respect to the first neighbor. However, for $k=2$ and $k=3$ we found 279 and 232 cases
of more extreme statistics. We can therefore
reject the null hypothesis of the same spatial distribution for G and NG systems at the $\sim$2.8\% 
and $\sim2.3$\% signficance levels, respectively. 
The difference between the spatial distributions of G and NG groups is reinforced by the
result of the two-dimensional KS test (e.g. Press et al. 1992; Fasano \& Franceschini 1987), which evaluates the
null hypothesis that a pair of data sets are drawn from the same distribution.
We found that this is rejected at the 95\% confidence level for all 2D data projections.

Combined, these results indicate that 
G groups probably inhabit denser regions, with higher densities of red galaxies, and are closer to each other (at least for $k=2$ and $k=3$) than NG groups -- a trend
that suggests an important environmental effect on structure formation through 
the UMaS volume.

\section{Discussion}

Superclusters are the largest structures identified in the universe. They
correspond to complex systems comprising a few up to several galaxy clusters and
groups experiencing long-term gravitational evolution. In this work, we studied
a galaxy sample $\sim50$ Mpc around an arbitrary center (the Abell cluster 1383) of the Ursa Majoris Supercluster.
We have addressed two questions: galaxy distribution around groups in the supercluster and the link
between the distribution of relaxed and nonrelaxed galaxy systems with respect to
the large-scale environment in which they are embedded.
We find that 68\% of galaxy groups are Gaussian and that all the
NG systems have substructures. 
This corresponds to a lower fraction of relaxed groups in comparison to
our previous studies (Ribeiro et al. 2010, 2011)\footnote{But consistent with Hou et al. (2009) for
CNOC2 data.}. This result cannot be  due to
the different criteria employed to split up G and NG groups, since we are being
more conservative to define nonrelaxed systems in the present work. Alternatively, this may result from 
the supercluster environment, which could be
favoring interactions and mergers between the groups. Actually, finding multimodality 
as the main cause of non-Gaussinity seems consistent with the idea of
ongoing processes in nonrelaxed systems  (e.g. Ribeiro et al. 2011; Hou et al. 2012; Einasto et al. 2012a).

 In this respect, an important issue relates to the possibility of also having NG groups as the result of 
spurious members in the velocity distributions. Wojtak et al. (2007) show that most statistical
methods (including the $3\sigma$-clipping procedure used in the present work) 
exclude 60\%-70\% of unbound galaxies on average. The probability of chance alignments with
the remaining interlopers was estimated with the following procedure: 

\begin{enumerate}
\item We assume that
all groups are 100\% reliable in the membership. 
\item We place all galaxies within them at the center,
making their velocity dispersions nil. 
\item And we rotate the cube around a random axis.
\item Then, we disperse the 
$los$ velocities according to a Gaussian distribution for each group with the original velocity
dispersion for the group. 
\item Finally, we apply an FoF code with the same parameters 
mentioned in Section 2, and analyze the Gaussianity of the resulting groups. 
\end{enumerate}

After 1,000
repetitions of this procedure, we found that $\sim$15\% of the mock groups, which are
Gaussian by construction, become NG, once additional members are included.
This means that about two NG groups in our sample would correspond to Gaussian systems with
spurious members in their velocity distributions. This fraction is too low to
signicantly modify the results we found in the previous sections, so most of
NG groups are probably portraits of multimodal systems.

 Even presuming this is so, Gaussian groups are still dominant across the supercluster, which suggests
the system is not young enough. Actually, Einasto et al. (2012b) show that superclusters
have two main morphological types, spiders and filaments. Clusters in superclusters of spider morphology have higher probabilities to have substructure and higher peculiar velocities of their main galaxies than clusters in superclusters of filament morphology. The Ursa Major supercluster is classified as filament-type in that work, which suggests it might be
dynamically evolved. In the same direction, we found that G groups inhabit denser regions,
with higher densities of both red and blue galaxies within 2.5 $h^{-1}$Mpc and are closer to each other
than NG groups, suggesting that relaxed systems may have formed and evolved earlier around high-density peaks, while
nonrelaxed systems may be growing more slowly in the peripheries of lower density peaks.
This shows the complex nature of superclusters so that further studies 
are needed to understand the role of the large-scale environment in galaxy group distribution and evolution.

\section{Acknowledgments}

We thank the referee Gary Mamon for very useful suggestions. We are thankful to Jason Pinkney
for making the substructure codes available.
ALBR acknowledges the support of the CNPq, grants 306870/2010-0 and 478753/2010-1. 
PAAL acknowledges the support of the FAPERJ, grant E-26/110.237/2010, and
CNPQ, grant 304692/2011-5.

\end{document}